\documentclass[twocolumn,showpacs]{revtex4}

\usepackage{amsmath}
\usepackage{graphicx}

\begin{document}

\title{Parametric Amplification of Nonlinear Response of Single Crystal
Niobium}
\author{Menachem I. Tsindlekht$^{\text{a}}$\thanks{%
email: mtsindl@vms.huji.ac.il}, Irena Shapiro$^{%
\text{b} }$, Moshe Gitterman$^{\text{b}}$}
\affiliation{$^{\text{a}}$The Racah Institute of Physics, The
Hebrew University of Jerusalem, 91904 Jerusalem, Israel}
\affiliation{$^{\text{b}}$Department of Physics, Bar Ilan
University , 52900 Ramat-Gan, Israel}
\date{\today}

\begin{abstract}
Giant enhancement of the nonlinear response of a single crystal Nb
sample, placed in {\it a pumping ac magnetic field}, has been
observed experimentally. The experimentally observed amplitude of
the output signal is about three orders of magnitude higher than
that seen without parametric pumping. The theoretical analysis
based on the extended double well potential model provides a
qualitative explanation of the experimental results as well as new
predictions of two bifurcations for specific values of the pumping
signal.

\end{abstract}
 \maketitle
\section{Introduction}
It was well established in the sixties  that nucleation of the
superconductivity in magnetic field parallel to the sample surface
begins when magnetic field $H$ becomes lower than surface
nucleation field $H_{c3}$ ($H_{c3}\approx 1.69 H_{c2}$) \cite{DG}.
The experiments at low frequencies reveal a nonlinear nature of
the ac response of the superconductors in surface superconducting
state \cite{ROLL,JOIN}. A critical state model for the
superconducting surface sheath was used for theoretical
explanation of the experimental findings \cite {ROLL,FINK}.
Recently our study of the nonlinear properties of single crystal
Nb showed that the critical state model hardly fits the new
experimental data and a simplified two level model was used for
the theoretical explanation of the experiment \cite{MT}.

 The above mentioned experiments in superconducting materials
 related to the nonlinear effects such as harmonic generation or
rectification. But there is another class of nonlinear effects -
parametric phenomena. Parametric phenomena are important in
fundamental physics and for applications. Parametric amplification
of the weak microwave signals in Josephson junctions were studied
in detail in thew seventies (see \cite{BP} and references
therein). Experimental observations of parametric amplification of
the weak microwave signals were possible because of the
nonlinearity of the Josephson junction \cite{BP}. To the best of
our knowledge parametric phenomena in surface superconducting
states of bulk superconductors has yet to be studied.

The present study is a continuation of our previous experimental
work and the theoretical analysis of the nonlinear dynamics in the
surface superconducting states of a single crystal Nb \cite{MT}.
In these experiments the dc and low-amplitude high-frequency ac
fields have been applied along the long axis of a sample. It was
found that in the steady-state regime the nonlinear effects exist
only in the surface superconducting phase. A unique aspect of this
system is that the non-linearity does not exist {\it a priory},
but rather appears only in the presence of an imposed dc field.

The new experiments have been performed in the same system,
subjected to an additional low-frequency excitation field, and the
amplitude of the output signal was measured as a function of the
amplitudes of the external fields including that of the ac
excitation field. In this experiment the nonlinearity of the
system plays a double role. Usual parametric experiments deal with
two signals. The system is excited by weak and strong (pumping)
signals. The frequency of the pumping signal is twice as large as
the frequency of the weak signal. Due to the nonlinearity of the
system, power from the pumping signal transfers to the weak signal
\cite{BP}. In our experiment, the system was excited by an
amplitude modulated ac field having some carrier and modulation
frequencies. The spectrum of the ac field does not contain the
modulation frequency. The modulation frequency in a system appears
when the sample is in a nonlinear state. Only for a nonlinear
system a pumping signal at a frequency twice as large as the
modulation frequency affects the nonlinear response. The
theoretical analysis, based on the extended double well potential
model of two surface states, provides a qualitative explanation of
the experimental data and predicts some new results which are
partially confirmed by experiment.

\section{Sample preparation and experimental technique}

A rectangular ($10\times 3\times 1$ mm$^{3}$) sample was cut by an
electric spark from a single crystal niobium bar which had been
fabricated by electron beam melting of high purity Nb precursor.
The [100] crystalline direction is parallel to the long axis of
the sample and the [010] direction is perpendicular to the sample
plane. After mechanical and chemical polishing
the crystal was annealed for one hour at T=1800 K under a vacuum of 10$%
^{-7}$ Torr. These procedures provided a high quality sample, as
confirmed by the high resistance ratio $R_{300K}/R_{10K}\approx
300$ and unperturbed critical temperature $T_{c}=9.15$ K. The
details of the employed Nb crystal characterization, including H-T
phase diagram, have been published elsewhere \cite{MT}.

The nonlinear response experiment was carried in a modified
commercial SQUID magnetometer, also used for measurements of the
dc magnetic properties. The Nb rod was placed in dc and ac
magnetic fields applied along the [100] direction. The excitation
ac field generated by a high frequency generator operating in a
constant current regime had the form of an amplitude modulated ac
field $h(t)=h_{0}(1+\alpha \cos \Omega t)\cos
\omega t$. Typical values of above parameters were $0<h_{0}<0.4$ Oe, $%
\alpha \approx 0.9$, $\omega /2\pi =3.2$ MHz, and $\Omega /2\pi
=1455$ Hz.  A parametric pumping field $%
h_p(t)=h_{2\Omega }\cos (2\Omega t)$ was achieved using a copper
solenoid of the commercial SQUID magnetometer system.

\begin{figure}
     \begin{center}
       \leavevmode
       \includegraphics[width=0.9\linewidth]{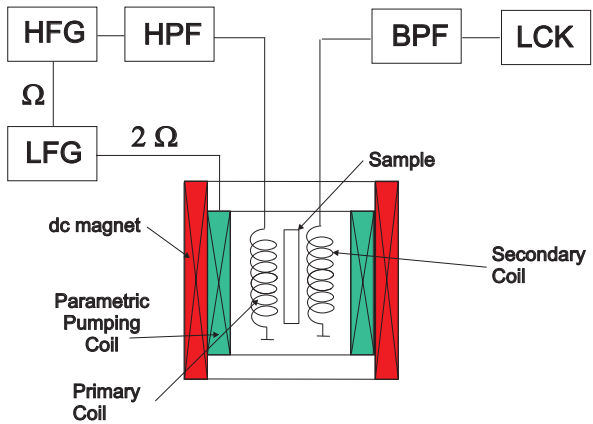}
       \bigskip
       \caption{(Color online) Block diagram of the experimental setup.
       HFC - high frequency generator, LFG - low frequency generator,
       HPF and BPF - high and band pass filters, LCK - lock-in amplifier.}
     \label{f-1a}
     \end{center}
     \end{figure}

A block diagram of the experimental setup is shown in Fig.
\ref{f-1a}. A 65-turn primary copper coil driven by a high
frequency generator (HP3325A) produces an amplitude modulated ac
field $h(t)$. Signals with frequencies $\Omega$ and $2\Omega$ were
generated by an Agilent 33250A model generator. The 1500 turn
secondary copper coil was wound directly on the primary coil. The
length of both coils was 15 mm. The voltage drop over the
secondary coil was measured by a lock-in amplifier (EG\&G 7265).
The modulation frequency $\Omega$ was used as a reference
frequency for the lock-in amplifier. The ac field amplitudes
$h_{0}$ and $h_{2\Omega}$ were measured by an additional small
probe coil wound under the primary coil. For the sake of clarity a
probe coil is not shown in Fig.~\ref{f-1a}.

One should note that the ratio between the pumping frequency and
the modulation frequency was equal exactly to 2 in our
experimental setup.

When the ac excitation is applied to a superconducting sample in
the nonlinear state, the magnetic moment of the sample oscillates
at the harmonics of the fundamental frequency $\omega $, at
frequencies $\omega \pm \Omega $, and at the frequency of
modulation $\Omega $ and its harmonics. The secondary coil
converts these magnetic moment oscillations into an ac voltage
signal. In the experiments we have measured, by means of lock-in
detection, the amplitude $A_{\Omega }$ of the signal at frequency
$\Omega $ as a function temperature, dc magnetic field $H$, ac
field amplitude $h_{0}$ and amplitude of parametric pumping $%
h_{2\Omega }$.

\section{Experimental results.}
\begin{figure}
     \begin{center}
       \leavevmode
       \includegraphics[width=0.9\linewidth]{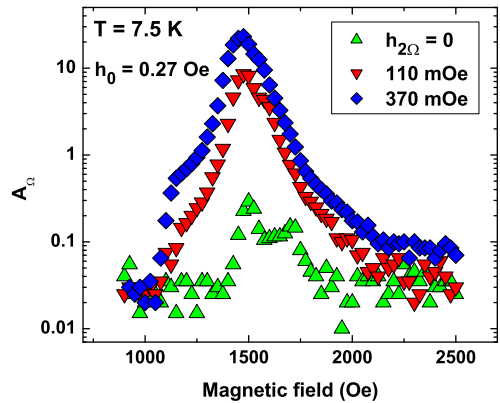}
       \bigskip
       \caption{(Color online) The amplitude of the output signal$A_{\Omega }(H)$ as %
        a function of the dc field $H$ at different pumping amplitudes $
h_{2\Omega }$.}
     \label{f-1}
     \end{center}
     \end{figure}

Here we restrict our consideration to the experimental data which
allow a direct theoretical explanation. Fig.~\ref{f-1} shows the
field dependence of the rectified signal $A_{\Omega }$ measured at
$T=7.5$ K for different amplitudes of $%
h_{2\Omega }$ compared with the data obtained in the absence of a
pumping field, $h_{2\Omega }=0$.

As evident from Fig.~\ref{f-1} the application of parametric
pumping leads to a giant enhancement of the rectified signal.
Under our experimental conditions the enhancement of $A_{\Omega}$
reaches two hundred times, the signal in absence of parametric
pumping. The present result confirms our resent observation
\cite{MT} that under stationary conditions the pronounced
rectified signal appears only in the surface superconducting
state.

\begin{figure}
     \begin{center}
       \leavevmode
       \includegraphics[width=0.9\linewidth]{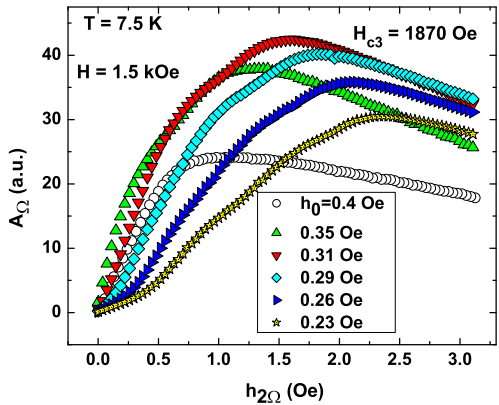}
       \bigskip
       \caption{(Color online) The amplitude of output signal $A_{\Omega }$ as %
        a function of pumping amplitude $h_{2\Omega }$.}
     \label{f-2}
     \end{center}
     \end{figure}

\begin{figure}
     \begin{center}
       \leavevmode
       \includegraphics[width=0.9\linewidth]{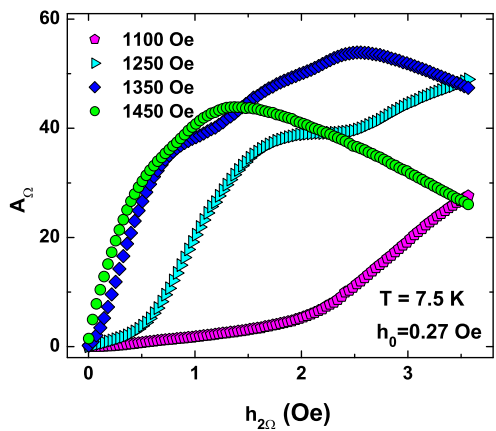}
       \bigskip
       \caption{(Color online) The amplitude of output signal $A_{\Omega }$ as %
        a function of pumping amplitude $h_{2\Omega }$ at different magnitude of magnetic field $H$.}
     \label{f-3}
     \end{center}
     \end{figure}

The measurements of $A_{\Omega}$ as a function of $h_{2\Omega}$
for constant $H$ and $h_0$ provides interesting information about
nonlinearity of the sample. In figures~\ref{f-2} and~\ref{f-3} we
present the amplitude of the output signal $A_{\Omega }$
as a function of the amplitude of the pumping field $%
h_{2\Omega }$ for different magnitudes of $H$ and amplitudes
$h_{0}$ of the dc and ac fields. Fig.~\ref{f-2} shows that
increasing the amplitude of excitation, $h_0$, does not always
lead to a straight enhancement of the amplitude of the nonlinear
response. Nonlinear systems with strong nonlinearity frequently
demonstrate this type of behavior.

The experimental data presented in Figs.~\ref{f-2},~\ref{f-3} are
very complicated. In order to straighten out this many-parameter
dependence a theoretical basis is needed which we will discuss in
the next section.

\section{ Theoretical analysis and discussions}

In the experiments described above the magnetic field $H$ %
is the large constant field, and $h\left( t\right) $ is a low
amplitude ac field. It turns out that for a given value of $H$
there are many superconducting states which can be labelled by the
number $n$ of fluxons pinned in this level, and each of them
corresponds to a local minima in the free energy. For the
theoretical description of the nonlinear response we use the
well-known Landau-Khalatnikov equation of motion for the magnetic
moment $M$ of the surface state
\begin{equation}
\frac{dM}{dt}=-\Gamma \frac{\partial G}{\partial M}  \label{1}
\end{equation}%
where $\Gamma $ is a phenomenological coefficient which hereafter
is assumed to be equal to unity, ($\Gamma =1$), and the total
Ginzburg-Landau free energy $G$ \ is composed of a series of
parabolic branches, $G=\sum G_{n}$, corresponding to\ the number
$n$ of the single flux quantum $\Phi _{0}$ in the state $G_{n}$
\cite{ROLL},
\begin{equation}
G_{n}=\frac{\pi R^{2}}{8}\left( M-M_{n}\right) ^{2}-\epsilon \left( 1-\frac{%
2M}{H_{c_{3}}}\right)  \label{2}
\end{equation}%
where $M_{n\text{ }}=n\Phi _{0}/\pi R^{2},$ and $\epsilon $ is the
energy to nucleate a fluxon.
\begin{figure}
     \begin{center}
       \leavevmode
       \includegraphics[width=0.35\linewidth]{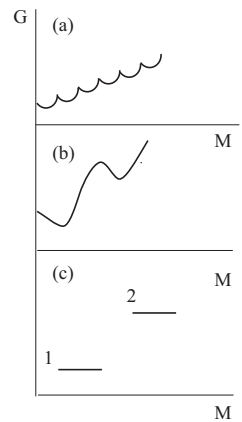}
       \bigskip
       \caption{(a) Schematic representation of the free energy of a
system as a function of the magnetic moment ; (b) a tilted double
well system ; (c) a two-level system.}
     \label{f-4}
     \end{center}
     \end{figure}

According to Eq.~(\ref{2}), the free energy as a function of an
external field is composed of a series of parabolic branches
(corresponding to different $n$) with centers located on an
ascending curve (Fig.~\ref{f-4}).

Instead of considering the free energy Eq.~(\ref{2}) we restrict
ourselves to the relevant initial and final states. In line with
this assumption let us replace the many-wells system shown in
Fig.~\ref{f-4}a, by a tilted double-well system (Fig.~\ref{f-4}b),
where the right minimum at $M=0$ corresponds to the surface
superconducting nucleation field $H_{C_{3}}$, and the left minimum
is related to some magnetic moment $M$ induced by an external
field $H.$ The distance between the minima is proportional to
$H_{C_{3}}-H,$

Hence, the adopted form of the free energy
\begin{equation}
\frac{4G}{\pi R^{2}}=-\frac{M^{2}}{2}+\frac{p^{2}M^{4}}{4}+\gamma
M \label{3}
\end{equation}%
contains two parameters%
\begin{equation}
p\approx \left( H_{C_{3}}-H\right) ^{-1}\text{ \ : \ }\gamma =\frac{%
2\epsilon }{H_{c_{3}}}  \label{4}
\end{equation}

Note that in our previous work~\cite{MT} we have used a model with
two discrete levels (Fig.~\ref{f-4}c). For discrete levels the
continuous dynamics, Eq.~(\ref{1}), is replaced by jumps between
the potential minima, and
the appropriate rate equation in the presence of the external ac field, \ %
$h\left( t\right)$, can be solved by using the adiabatic
approximation. In such a manner we have accounted for the main
experimental results such as a set of maxima of the output signal
as a function of $H$ for different fixed ac strength
$h_{0}$~\cite{MT}.

In the experiments analyzed in the previous publication~\cite{MT}
the external ac field was the sum of three monochromatic waves
\begin{equation}
\begin{array}{c}
h(t)=h_{0}\cos (\omega t)+0.45h_{0}[\cos (\omega +\Omega )t\\+\cos
(\omega -\Omega )t]  \label{5}
\end{array}
\end{equation}%
one with a frequency $\omega $, and two with satellite frequencies
$\omega \pm \Omega $, where $\Omega \ll \omega $. This
multifrequency excitation of a non-linear system allows one to
detect a signal at low frequency $\Omega $.

Unlike the previous work, in the new experiments described in the
previous sections , an additional pumping field
$h_{p}(t)=h_{2\Omega }cos(2\Omega t)$
was applied to the system. Substituting this field together with Eqs.~(\ref{2}%
-\ref{4}) into Eq.~(\ref{1}) one gets
\begin{equation}
\begin{array}{c}
\frac{4}{\pi R^{2}\Gamma }\frac{dM}{dt}=M-(\frac{1}{H_{c3}-H})^{2}M^{3}+2%
\frac{\epsilon }{H_{c3}}\\+h_{0}\cos \omega t +0.45h_{0}[\cos
(\omega +\Omega )t\\+\cos (\omega -\Omega )t]+h_{2\Omega }\cos
2\Omega t%
\end{array}
\label{6}
\end{equation}

Since $\Omega \ll \omega $, one can decompose $M\left( t\right) $
into a sum of two terms as follows \cite{GIT}

\begin{equation}
\begin{array}{c}
M\left( t\right) =x\left( t\right) +\frac{h_{0}}{\omega }\sin
\left( \omega t\right) +\\\frac{0.45h_{0}}{\left( \omega +\Omega
\right) }\sin \left[ \left( \omega +\Omega \right)t\right]+ \\
\frac{0.45h_{0}}{2\left( \omega -\Omega \right) }\sin \left[
\left( \omega -\Omega \right) t\right]
\end{array}
  \label{7}
\end{equation}

The first term on the right-hand side of Eq.~(\ref{7}) will be
assumed to vary significantly only over times of the order of $t$,
while the other terms vary rapidly. Substituting~(\ref{7})
into~(\ref{6}) one can perform an averaging over a single cycle
time of $\sin \left(\omega t\right)$. All odd powers of $\sin
\left( \omega t\right) $ vanish under the average while the $\sin
^{2}\left( \omega t\right) $ term will give $\frac{1}{2}.$
Finally, one obtains the following equation for $A\left( t\right)
,$ the mean value of $M\left( t\right) $ during the oscillation
period, $A\left(
t\right) =$ $<M\left( t\right) >,$%
\begin{equation}
\begin{array}{c}
\frac{dA}{d\tau }-A\left[ a-b\cos \left( \Omega t\right) -c\cos
\left( 2\Omega t\right) \right]\\ +p^{2}A^{3}+\beta=
\frac{h_{2\Omega }}{h_{0}}
 \cos \left( 2\Omega t\right)
\end{array}
\label{8}
\end{equation}
where we have introduced the dimensionless parameters%
\begin{equation}
\begin{array}{c}
A=\frac{M}{h_{0}}\text{;\ }\tau =\frac{4t}{\pi R^{2}}\text{;\
}p^{2}=\left( \frac{h_{0}}{H_{c3}-H}\right) ^{2}\text{; }\\\beta
=\frac{\gamma }{h_{0}}
\text{\ }a=1-\frac{9.9p^{2}}{4\left( \omega \pi R^{2}\right) ^{2}}\text{; }\\b=%
\frac{3.3p^{2}}{\left( \omega \pi R^{2}\Gamma \right) ^{2}}\text{ ; \ }c=%
\frac{3.3p^{2}}{4\left( \omega \pi R^{2}\right) ^{2}}%
\end{array}
\label{9}
\end{equation}

It is clear now that for dimensionless amplitudes of the output
and the input signal, $\frac{M}{h_{0}}$ and $\frac{h_{2\Omega
}}{h_{0}}$, the main
parameter which defines the input-output relations is $\left( \frac{h_{0}}{%
H_{c3}-H}\right) $ while the phenomenological parameter $\beta $
is less important.

\begin{figure}
       \includegraphics[width=0.9\linewidth]{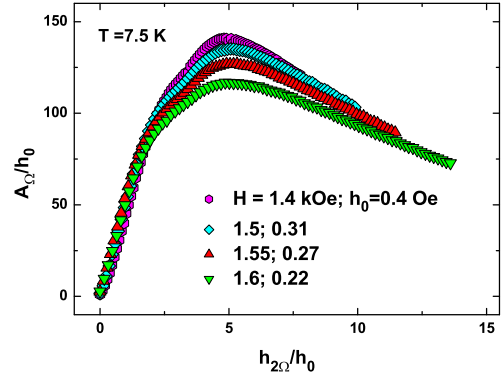}
       \caption{(Color online) Experimental data redrawn in the dimensionless
amplitudes of the output and the input signals, $\frac{A_{\Omega}}{h_{0}}$ and $\frac{h_{2\Omega }}{h_{0}%
}$ for the constant value of the dimensionless parameter $\frac{h_{0}}{%
H_{c3}-H}$.}
     \label{f-5}
     \end{figure}

Based on the theory described above we may now redraw the
experimental data shown in Figs.~\ref{f-2} and~\ref{f-3}. In order
to straighten out these experimental data
one has to use the scaled dimensionless coordinates $\frac{A}{h_{0}}$ and $%
\frac{h_{2\Omega }}{h_{0}},$ and grouping together the
experimental points \ corresponding to the same value of the
parameter $\frac{h_{0}}{H_{c3}-H}.$ These regrouping data from
Figs.~\ref{f-2} and~\ref{f-3} are show in Fig.~\ref{f-5}.

\begin{figure}
    \begin{center}
       \leavevmode
\includegraphics[width=0.9\linewidth]{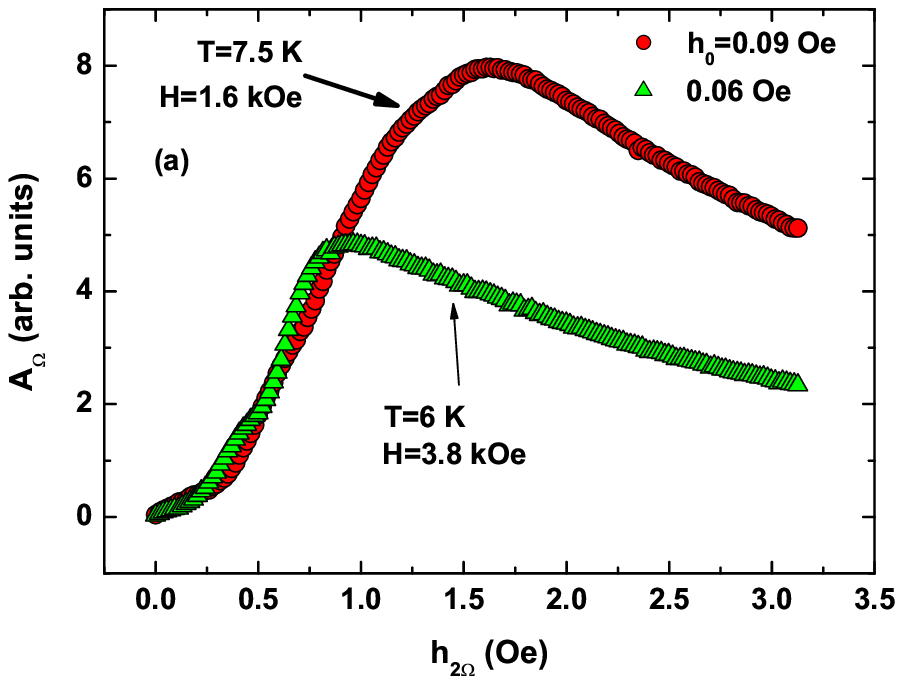}
\includegraphics[width=0.9\linewidth]{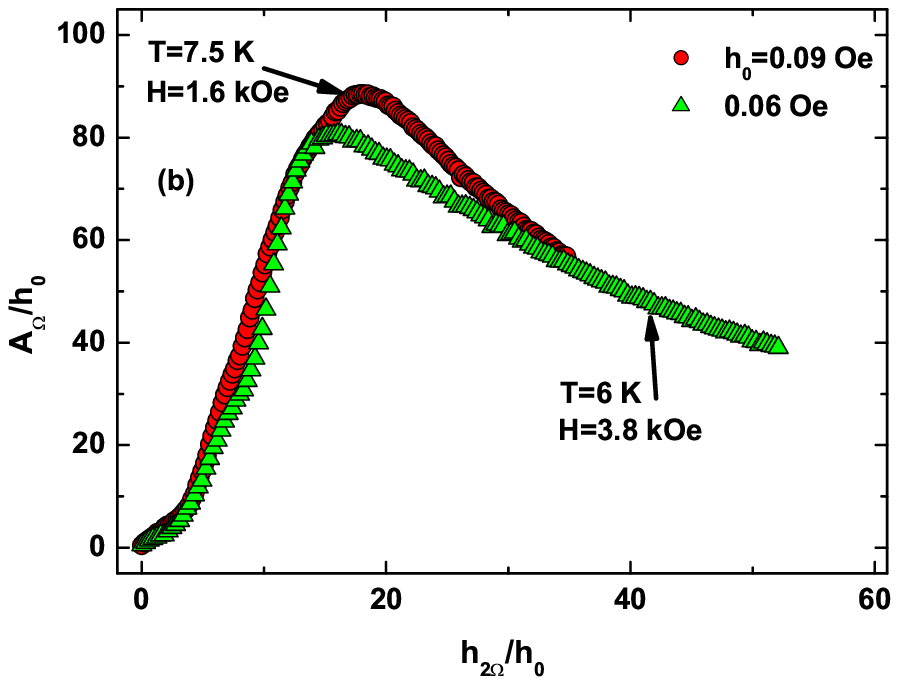}
\caption{(Color online) (a) $A_{\Omega}(h_{2\Omega}$ dependencies
for $T=7.5$ K
 and $T=6$ K for the constant value of the dimensionless parameter $\frac{h_{0}}
 {H_{c3}-H}$.
  (b) The same experimental data redrawn in the dimensionless
amplitudes of the output signal and the parametric pumping amplitude, $\frac{A_{\Omega}}{h_{0}}$ and $\frac{h_{2\Omega }}{h_{0}%
}$ }
     \label{f-6a}
     \end{center}
     \end{figure}

 As one can see from
this figure the experimental data are now well ordered as they
should be according to Eq.~(\ref{8}). The small differences
between different curves might be attributable to the different
values of parameter $\beta $. The regrouping procedure was used
for the evaluation of $H_{c3}$ from the experimental data as well.
The universal behavior $A_{\Omega}(h_{2\Omega})$ occurs possible
for different temperatures. Fig. \ref{f-6a} demonstrates this type
of universal behavior for $A_{\Omega}(h_{2\Omega})$ in
dimensionless coordinates at $T=7.5$ K and $T=6$ K .

\begin{figure}
     \begin{center}
       \leavevmode
       \includegraphics[width=0.9\linewidth]{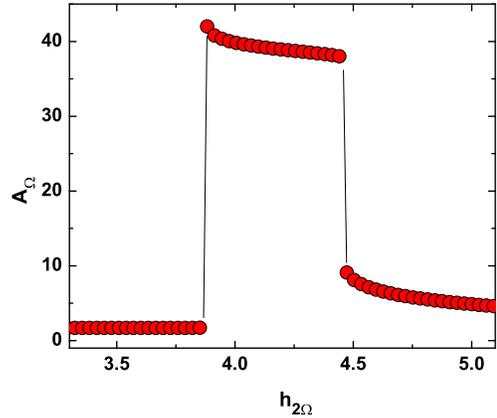}
       \caption{(Color online) Two bifurcations of the amplitude of the output signal
$A_{\Omega }$ as a function of the amplitude $h_{2\Omega }$ of the
pumping signal.}
     \label{f-6}
     \end{center}
     \end{figure}

The universal form of the curves in Figs.~\ref{f-5},\ref{f-6a}
suggests the validity of our theory. Eq. (\ref{8}) describes the
overdamped Duffing oscillator with parametric and forced periodic
excitations. Although an extensive literature is devoted to an
investigation of the Duffing oscillator(\cite{NA}), the periodic
excitations were analyzed only for the underdamped case
(\cite{FX}). We have performed comprehensive numerical
integrations of Eq.~(\ref{8}). A
typical result of these calculations for some values of the parameter $\frac{h_{0}%
}{H_{c3}-H}$ is shown in Fig.~\ref{f-6}.

Several conclusions can be reached from the analysis of this and
similar graphs :

1. Over a wide range of the parameter $\frac{h_{0}}{H_{c3}-H},$
there are two bifurcations at small (\textquotedblright
left\textquotedblright ) and large ("right" ) values of the
amplitudes of the pumping field.

2. With a decrease of $(\frac{h_{0}}{H_{c3}-H})^2$, the
coefficient which multiplies the nonlinear term, the right
bifurcation occurs at a larger amplitude of the pumping signal
$,$and the jump becomes larger.

3. With a decrease of $b$ (and $c$) which describe the parametric
excitation, the left bifurcation occurs at a somewhat larger
pumping amplitude, and the jump becomes slightly larger.

\begin{figure}
     \begin{center}
       \leavevmode
       \includegraphics[width=0.9\linewidth]{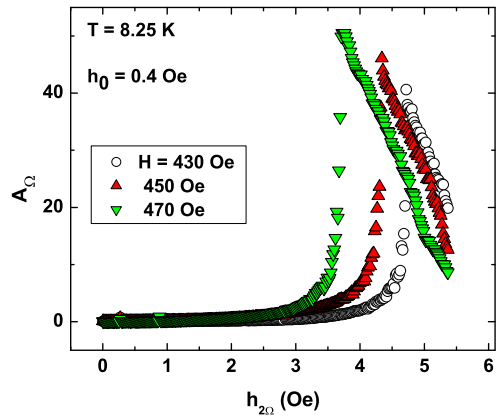}
       \caption{(Color online) Experimental data describing the dependence of the
amplitude of the output signal $A_{\Omega }$ as a function of the
amplitude of the pumping signal $h_{2\Omega }$ for different
values of $H$.}
     \label{f-7}
     \end{center}
     \end{figure}

\begin{figure}
     \begin{center}
       \leavevmode
       \includegraphics[width=0.9\linewidth]{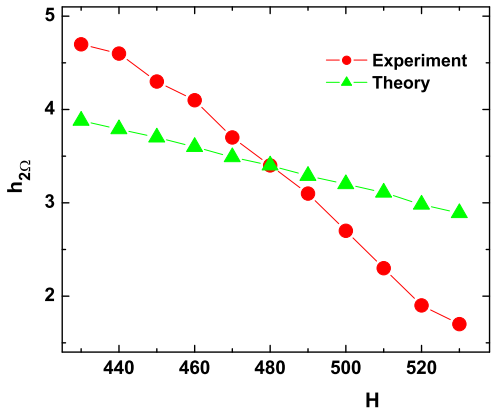}
       \caption{$h_{2\Omega }$ of maxima
obtained from experimental graphs Fig.~\ref{f-7} and from the
theoretical analysis of Eq.~\ref{8}.}
     \label{f-8}
     \end{center}
     \end{figure}

The currently available experimental data show at least
qualitative agreement with the suggested theory:

a) Comparison of the experimental graphs in figures~\ref{f-2}\\
and~\ref{f-3}, and their scaling treatment shown in
Figs.~\ref{f-5},\ref{f-6a} clearly proves that the theoretical
model addresses the main features of the experimental data.

b) The giant increase of the output signal in the presence of the
pumping field shown in Fig.~\ref{f-1}, is present in our
theoretical model as well. For instance, for $p=2\ast 10^{-2}$\ \
, equation~(\ref{8}) shows a two order of
magnitude increase in $A_{\Omega }$ in going from $h_{2\Omega }=0$ to $%
h_{2\Omega }=50.$

c) The presence of the left bifurcation can be seen from
experimental graph~\ref{f-7} which shows $A_{\Omega }$ as a
function of \ h$_{2\Omega }$ for different values of $H$.
The values of $h_{2\Omega }$ at maxima are shown in Fig. \ref{f-8}
for different $H$. In the same figure we show also the same
quantities obtained from Eq. (\ref{8}). Since in the theoretical
analysis we omit some constant coefficients ($\Gamma =1,p\approx
\left( H_{C_{3}}-H\right) ^{-1}$, etc.), in the comparison of
theoretical results with experimental data one has to set one of
the points on the theoretical curve with one on the experimental
curve, which has been done in Fig. \ref{f-8}.

Both curves are almost straight lines with the correct sign of the
slope, i.e., we conclude that our theoretical model gives at least
semi-quantitative agreement with experiment. However, one cannot
see the right bifurcation on the experimental graph 8 which is,
probably, connected with some ignored  factor in our theory which
smears this bifurcation. Another possible explanation lies in the
fact that in order to see in the experiment the right
bifuracation, one needs the higher amplitudes of the pumping
signal which cannot be achieved with the existing experimental set
up.

\section{Summary}
We have performed an extensive experimental study of the
non-linear response in  surface superconducting states of single
crystal niobium. Application of a pumping ac magnetic field
results in a giant (about three order of magnitude) increase of
the amplitude of the output signal. The theoretical analysis,
based on the extend double well potential model, provides a
qualitative explanation of the experimental results as well as
predicting a new phenomenon, the abrupt change (bifurcation) of
the nonlinear response in ac driven superconductors. The latter
effect, in its turn was partially supported by experiment.

\bigskip
\section{Acknowledgments}

The work at the Hebrew University was supported by the Klatchky
foundation. The authors are deeply grateful to Dr. S.I. Bozhko and
Dr. A.M. Ionov (ISSP RAS, Chernogolovka, Moscow district, Russia)
for the sample preparation. We thank Professors I. Felner, B.Ya.
Shapiro, Dr. V. Genkin and Dr. G. Leviev
 for useful discussions. M.I.T. wishes to thank G.
Greenwald for the providing nonstandard electronic devices.

\end{document}